\documentclass[english,aps,prl,twocolumn,superscriptaddress,showpacs,tightenlines]{revtex4}
\usepackage[T1]{fontenc}
\usepackage[latin9]{inputenc}
\usepackage{color}
\usepackage{babel}
\usepackage{amsmath}
\usepackage{amssymb}
\usepackage{graphicx}
\usepackage[unicode=true,pdfusetitle,
 bookmarks=true,bookmarksnumbered=false,bookmarksopen=false,
 breaklinks=false,pdfborder={0 0 1},backref=false,colorlinks=true]
 {hyperref}
\hypersetup{
 citecolor=blue}
\usepackage{breakurl}

\makeatletter
\@ifundefined{textcolor}{}
{%
 \definecolor{BLACK}{gray}{0}
 \definecolor{WHITE}{gray}{1}
 \definecolor{RED}{rgb}{1,0,0}
 \definecolor{GREEN}{rgb}{0,1,0}
 \definecolor{BLUE}{rgb}{0,0,1}
 \definecolor{CYAN}{cmyk}{1,0,0,0}
 \definecolor{MAGENTA}{cmyk}{0,1,0,0}
 \definecolor{YELLOW}{cmyk}{0,0,1,0}
}

\newcommand*{\ket}[1]{|{#1}\rangle}

\newcommand*{\bra}[1]{\langle{#1}|}

\usepackage{soul}

\makeatother

\begin{document}

\title{Entropy Production of Open Quantum System in Multi-Bath Environment}

\author{Cheng-Yun Cai}

\affiliation{State Key Laboratory of Theoretical Physics, Institute of Theoretical
Physics and University of the Chinese Academy of Sciences, Beijing
100190, People's Republic of China}

\affiliation{Synergetic Innovation Center of Quantum Information and Quantum Physics,
University of Science and Technology of China, Hefei, Anhui 230026,
China}

\author{Sheng-Wen Li}

\affiliation{Beijing Computational Science Research Center, Beijing 100084, China}

\affiliation{Synergetic Innovation Center of Quantum Information and Quantum Physics,
University of Science and Technology of China, Hefei, Anhui 230026,
China}

\author{Xu-Feng Liu}

\affiliation{Department of Mathematics, Peking University, Beijing 100871, China}

\author{C. P. Sun}

\email{cpsun@csrc.ac.cn}

\homepage{http://www.csrc.ac.cn/suncp/}

\selectlanguage{english}%

\affiliation{Beijing Computational Science Research Center, Beijing 100084, China}

\affiliation{Synergetic Innovation Center of Quantum Information and Quantum Physics,
University of Science and Technology of China, Hefei, Anhui 230026,
China}
\begin{abstract}
We study the entropy production of an open quantum system surrounded
by a complex environment consisting of several heat baths at different
temperatures. The detailed balance is elaborated in view of the distinguishable
channels provided by the couplings to different heat baths, and a
refined entropy production rate is derived accordingly. It is demonstrated
that the entropy production rates can characterize the quantum statistical
property of the baths: the bosonic and fermionic baths display different
behaviors in the high-temperature limit while they have the same asymptotic
behavior at low temperature.
\end{abstract}


\maketitle
\emph{Introduction.}---The concept of entropy plays an important role
in our understanding of complex physical systems. For a closed system
its entropy will never decrease if the unitarity condition is satisfied
whether the system enjoys time reversal invariance or not \cite{Yang 1980,Thomsen 1953}.
For an open system contacting with its environment entropy production
is a pivotal concept. The so called entropy production rate (EPR)
is usually regarded as a signature of the irreversibility associated
with such a system \cite{Prigogine1977}. In fact, EPR is a proper
physical quantity tagging the steady state of an open system. A lot
of open systems can be well studied in the framework of time-homogeneous
Markov process. The steady states of such systems fall into two categories:
equilibrium steady states and non-equilibrium ones. In a non-equilibrium
steady state the detailed balance is broken, or equivalently, the
EPR does not vanish. Thus one can say that the system is in an equilibrium
steady state if and only if the accompanying EPR is zero.

It should be noted that from purely mathematical point of view the
reversibility of a time-homogeneous Markov process and the detailed
balance condition has been thorough studied by Kolmogorov \cite{Kolmogorov1936}.

For a Markov process determined by the Pauli master equation $\dot{p}_{i}=\sum_{j}p_{j}L_{ji}-p_{i}L_{ij}$,
the breaking of detailed balance is quantitively characterized by
a non-vanishing EPR \cite{Schnakenberg1976,M Qian2012,M Qian 2004,Spohn1978}
\begin{equation}
R=\frac{1}{2}\sum_{ij}(p_{i}L_{ij}-p_{j}L_{ji})\ln\frac{p_{i}L_{ij}}{p_{j}L_{ji}}.\label{EP_classical}
\end{equation}
Here, $p_{j}$ is the probability of the open system with which the
open system appears in state-$j$ and $L_{ij}$ is the transition
rate from state-$i$ to state-$j$. If the detailed balance is satisfied,
i.e., $p_{i}L_{ij}=p_{j}L_{ji},$ there is no entropy production when
the system reaches the steady state. This observation has been the
starting point of a fruitful study of some non-equilibrium biochemical
reactions (see Ref.\,\cite{M Qian 2000} and references therein). 

\begin{figure}
\begin{centering}
\includegraphics[scale=0.5]{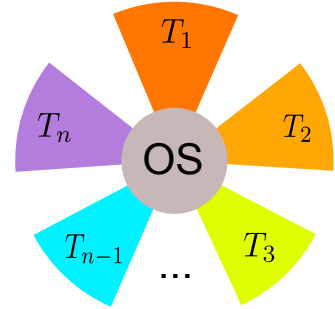}
\par\end{centering}

\protect\caption{\label{fig1}(color online). A complicated non-equilibrium process
of an open system (OS) contacting with $n$ heat baths, where the
temperature of the $i$-th heat bath is $T_{i}$.}
\end{figure}

The above formula of EPR is valid for a classical system contacting
with a single canonical heat bath. For the system in a complex environment
consisting of two or more heat baths at different temperatures (see
Fig.\,\ref{fig1}), which allows more complicated non-equilibrium
processes \cite{Esposito 2007}, the formula should be generalized.
In fact, if the conventional EPR formula (\ref{EP_classical}) were
applied in this case, we would obtain a vanishing EPR whenever the
system reaches the steady state. This contradicts the intuitional
physical picture. We notice that the non-equilibrium processes induced
by the multi-bath environment emerge in many practical systems, such
as an opto-mechanical system with quantum cavity field coupled to
two moving cavity wells at different temperatures \cite{H Ian2008},
and a quantum dot system gated by two electrodes \cite{Gurvitz1998,Esposito 2009}
at different temperatures.

In this letter, we first derive a quantum EPR formula for the above
mentioned multi-bath case from the master equation of the open quantum
system. When the quantum coherence is neglected, our general result
reduces to the multi-channel expression of EPR as given in Ref.\,\cite{Esposito 2007}.
Then we show that the EPRs of non-equilibrium systems with bosonic
and fermionic baths have a similar behavior at low temperature but
behave quite differently in the high temperature region. This leads
to the conclusion: non-equilibrium process is important as it can
characteristically reflect the quantum statistical property of the
bath.

\emph{Entropy Production Rate for Open Quantum Systems.}---The Markovian
evolution of an open quantum system in contact with its environment
is described by a dynamical semigroup $\Lambda_{t}$. To be precise,
we have $\rho(t)=\Lambda_{t}\rho(0)$ and the density matrix $\rho(t)$
satisfies the master equation $\dot{\rho}={\cal L}[\rho]$ with Lindblad
super-operator ${\cal L}$ (we always work in the interaction picture
hereafter)\,\cite{Lindblad 1976}. For a heat bath at temperature
$T$, it has been shown that the EPR can be formulated by the relative
entropy as $R^{(0)}=-\frac{\mathrm{d}}{\mathrm{d}t}S(\rho||\rho^{\mathrm{th}})$\cite{Spohn1978,Breuer2002}.
Here, the reference state $\rho^{\mathrm{th}}=\exp[-\beta H]/Z$ is
the steady state given by ${\cal L}[\rho^{\mathrm{th}}]=0$, and it
is the thermal equilibrium state at the temperature $T=\beta^{-1}$.
The EPR formula is decomposed into two terms, one of which is $-\frac{\mathrm{d}}{\mathrm{d}t}\mathrm{tr}[\rho\ln\rho]$,
representing the entropy changing rate of the open system itselfand
the other of which takes the form: 
\begin{equation}
\frac{\mathrm{d}}{\mathrm{d}t}\mathrm{tr}(\rho\ln\rho^{\mathrm{th}})=\frac{\mathrm{d}}{\mathrm{d}t}\mathrm{tr}[\rho(-\beta H)]=-\frac{1}{T}\frac{\mathrm{d}\left\langle H\right\rangle }{\mathrm{d}t}.
\end{equation}
Since $-\frac{\mathrm{d}}{\mathrm{d}t}\left\langle H\right\rangle :=\dot{Q}$
is the rate of the heat dissipating into the heat bath \cite{HT Quan2005},
this term describes the entropy changing rate of the heat bath. Thus
this EPR formula actually counts the total entropy changing rate of
both the open system and its environment.

For an open quantum system interacting with $N$ reservoirs (Fig.\,\ref{fig1}),
the master equation assumes the form $\dot{\rho}=\sum_{l=1}^{N}{\cal L}_{l}[\rho]$,
where ${\cal L}_{l}$ is the Lindblad super-operator corresponding
to the $l$-th reservoir. In time interval $\mathrm{d}t$, the energy
dissipating into reservoir-$l$ is $\mathrm{\mkern3mu\mathchar'26\mkern-12mu d}Q_{l}:=-\mathrm{tr}\big[{\cal L}_{l}[\rho]H\big]\mathrm{d}t$
\cite{HT Quan2005}, thus the entropy changing rate through the $l$-th
reservoir is
\begin{equation}
\frac{\dot{Q}_{l}}{T_{l}}=-\frac{1}{T_{l}}\mathrm{tr}\big[{\cal L}_{l}[\rho]H\big],
\end{equation}
where $\rho_{l}^{\mathrm{th}}:=\exp[-\beta_{l}H]/Z_{l}$ is the thermal
state of reservoir-$l$ with temperature $\beta_{l}^{-1}$. Then the
entropy production rate for a non-equilibrium quantum system is obtained
as $R=\frac{\mathrm{d}}{\mathrm{d}t}\mathrm{tr}[\rho\ln\rho]+\sum_{l}\dot{Q}_{l}/T_{l}$,
i.e,
\begin{equation}
R=-\sum_{l=1}^{N}\mathrm{tr}\big[{\cal L}_{l}[\rho](\ln\rho-\ln\rho_{l}^{\mathrm{th}})\big],\label{epr_quantum_multi-bath}
\end{equation}
which represents the total entropy changing rate of the open system
and its multi-bath environment. It has been proved that the quantity
$R_{l}:=-\mathrm{tr}\big[{\cal L}_{l}[\rho](\ln\rho-\ln\rho_{l}^{\mathrm{th}})\big]$
is non-negative \cite{Spohn1978}, so we always have $R=\sum_{l}R_{l}\geqslant0$.

The dynamics of the quantum coherence is usually decoupled from that
of the populations (when we say quantum coherence, we mean the effect
contributed from the off-diagonal terms $\langle i|\rho|j\rangle$
of $\rho$ in the eigen energy representation) \cite{Breuer2002}.
When the quantum coherence is neglected (the effect of the quantum
coherence will be studied later), the Lindblad equation reduces to
Pauli master equation, and the above Eq.\,(\ref{epr_quantum_multi-bath})
reduces to
\begin{equation}
R\simeq-\sum_{l=1}^{N}\sum_{i}\bra{i}{\cal L}_{l}[\rho]\ket{i}\bra{i}(\ln\rho-\ln\rho_{l}^{\mathrm{th}})\ket{i}.
\end{equation}
Replacing $\bra{i}{\cal L}_{l}[\rho]\ket{i}$ by $\sum\limits _{j}(p_{j}L_{ji}^{(l)}-p_{i}L_{ij}^{(l)})$,
we obtain
\[
R=\frac{1}{2}\sum_{l=1}^{N}\sum_{i,j}(p_{j}L_{ji}^{(l)}-p_{i}L_{ij}^{(l)})(\ln\frac{p_{i}}{p_{j}}+\frac{E_{i}-E_{j}}{T_{l}})
\]
 Here $L_{ij}^{(l)}$ is the transition rate from state-$i$ to state-$j$
resulted from the coupling to the $l$-th heat bath, and we have used
the fact that $\bra{i}\rho_{l}^{\mathrm{th}}\ket{i}=\exp(-E_{i}/T_{l})/Z_{l}$.
Considering the microscopic reversibility condition $L_{ij}^{(l)}/L_{ji}^{(l)}=\exp[(E_{i}-E_{j})/T_{l}],$
we then have 
\begin{equation}
R=\frac{1}{2}\sum_{i\not=j}\sum_{l=1}^{N}(p_{i}L_{ij}^{(l)}-p_{j}L_{ji}^{(l)})\ln\frac{p_{i}L_{ij}^{(l)}}{p_{j}L_{ji}^{(l)}},\label{REPR}
\end{equation}
a refined entropy production rate. This is the same as Eq.\,(10)
in Ref.\,\cite{Esposito 2007}. If the quantum coherence is taken
into account, there would be extra entropy production.

\begin{figure}

\begin{centering}
\includegraphics[width=5cm]{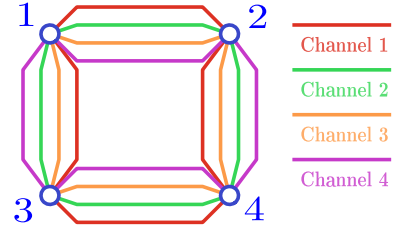}
\par\end{centering}

\protect\caption{\label{fig3-1}(color online). The transition diagram of a four-level
system contacting with four heat baths. We use different colors to
distinguish the channels corresponding to different heat baths. }

\end{figure}

Note that in the above refined entropy production rate (REPR) transitions
caused by different heat baths (see Fig.\,\ref{fig3-1}) are treated
separately while in the spirit of the conventional EPR Eq.\,(\ref{EP_classical})
they should be merged. This essential difference naturally leads to
different understandings of equilibrium state.

\emph{Elaborate Detailed Balance and Time Reversibility.}---As we
have argued above, if the environment is composed of two or more heat
baths, the REPR~(\ref{REPR}), instead of the conventional EPR Eq.\,(\ref{EP_classical}),
should be used to investigate the entroy production problem. Then,
the condition for zero EPR is refined as
\begin{equation}
p_{i}L_{ij}^{(l)}=p_{j}L_{ji}^{(l)}.\label{EDB}
\end{equation}
This condition is subtler than the detailed balance condition $p_{i}L_{ij}=p_{j}L_{ji}$
and justifies the name of elaborate detailed balance (EDB). The quantity
$L_{ij}^{(l)}$ can be viewed as the transition rate from state-$i$
to state-$j$ through the $l$-th channel. From this point of view,
the EDB requires not only the balance of transitions between any two
states of the system, but also the balance of each transition channel.
This property has been suggested by Lewis as a criterion for equilibrium
\cite{Lewis 1925}.

\begin{figure}
\begin{centering}
\includegraphics[width=6cm]{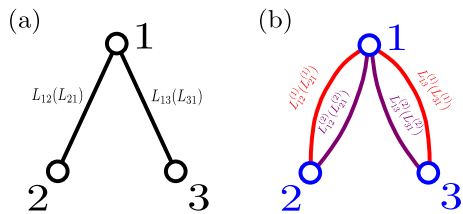}
\par\end{centering}

\protect\caption{\label{fig3}(color online). The transition diagram of a $\Lambda$-type
system contacting with two heat baths. (a) A rough description of
the system's possible transitions, $1\leftrightarrow2$ and $1\leftrightarrow3$.
(b) Each transition in (a) may be further divided into two channels,
each of which corresponds to a transition caused by a certain heat
bath.}
\end{figure}

Let us probe further the concept of transition channel with the case
of a $\Lambda$-type system contacting with two heat baths. We consider
the transition from state-1 to state-2, which releases energy of amount
$\omega_{12}$ to its environment. If there appears an energy increase
of $\omega_{12}$ in the first heat bath, then it can be judged that
this transition results from the first heat bath. Thus it is physically
justified to understand the transition as realized through two distinguishable
channels, each of which corresponds to a transition caused by a certain
heat bath. As a consequence of the validity of the concept of distinguishable
channels it can be argued that a transition chain like $\cdots1\rightarrow2\rightarrow1\rightarrow3\cdots$
is not a complete description. A complete description should be something
like $\cdots1\stackrel{(1)}{\longrightarrow}2\stackrel{(2)}{\longrightarrow}1\stackrel{(2)}{\longrightarrow}3\cdots$,
where $1\stackrel{(1)}{\longrightarrow}2$ denotes a transition from
state-1 to state-2 via the first channel and $1\stackrel{(2)}{\longrightarrow}3$
denotes a transition from state-1 to state-3 via the second channel.
All the possible channels together make up a transition diagram as
shown in Fig.\,\ref{fig3}b. 

Under certain conditions the equivalence between the conventional
detailed balance and the time reversibility of a Markov process has
been proved by Kolmogorov \cite{Kolmogorov1936}. As there is no generic
way to model the evolution of the open system with distinguishable
transition channels as a mathematically well defined Markov process,
Kolmogorov's result is not immediately applicable to the open system
with two or more baths. Nevertheless, from physical intuition, one
may well expect that the time reversibility of such systems requires
that the likelihood of transitions $1\stackrel{(2)}{\longrightarrow}3$
be the same in the forward process and backward process: $p_{1}L_{13}^{(2)}=p_{3}L_{31}^{(2)}$.
Furthermore, it is also intuitively correct that the EDB will guarantee
the time reversibility of the dynamics of such systems.

Since the open systems considered here are mesoscopic or microscopic,
the evolution should be subjected to quantum dynamics. There exist
mainly two kinds of quantum effects to be considered in the entropy
production, related to quantum statistics and quantum coherence respectively.

\emph{Quantum Statistical Effect on Entropy Production Rate.}--- Let
us first consider the factor of quantum statistics. Bosonic and fermionic
environments are fundamental in the study of quantum open systems.
Bosonic heat baths usually appear in opto-mechanical systems while
fermionic ones are common in the study of quantum dots. What is the
essential difference between these two basic kinds of environments?
In this section we try to answer this question from entropy production
point of view. Our starting point is the REPR formula (\ref{REPR}),
applied to a two-level system.

For a two-level system contacting with two heat baths, the REPR {[}Eq.\,(\ref{REPR}){]}
for the steady state is
\begin{equation}
R=\frac{\Omega}{L_{12}+L_{21}}\cdot\frac{T_{1}-T_{2}}{T_{1}T_{2}}(L_{21}^{(2)}L_{12}^{(1)}-L_{12}^{(2)}L_{21}^{(1)}).\label{eq:13}
\end{equation}
Here, we have adopted the labeling: state-$1$ and state-$2$ denote
the excited state and the ground state respectively. Different kinds
of environments will lead to different forms of transition rates.
Specifically, the transition rates caused by the$l$-th heat bath
have the following forms \cite{Breuer2002},
\begin{eqnarray}
L_{12}^{(l)} & = & \gamma_{l}(1\pm N(\beta_{l})),\label{eq:14}\\
L_{21}^{(l)} & = & \gamma_{l}N(\beta_{l}),\label{eq:15}
\end{eqnarray}
where `$+$' and `$-$' correspond to the bosonic and fermionic cases
respectively. $N(\beta)=1/(\exp(\beta\Omega)\mp1)$ is the distribution
function, and $\gamma_{l}$ is the coupling strength between the system
and the $l$-th heat bath. One can verify that these transition rates
satisfy the microscopic reversibility $L_{12}^{(l)}=\exp(-\beta_{l}\Omega)L_{21}^{(l)}$. 

\begin{figure}
\begin{centering}
\includegraphics[width=8cm]{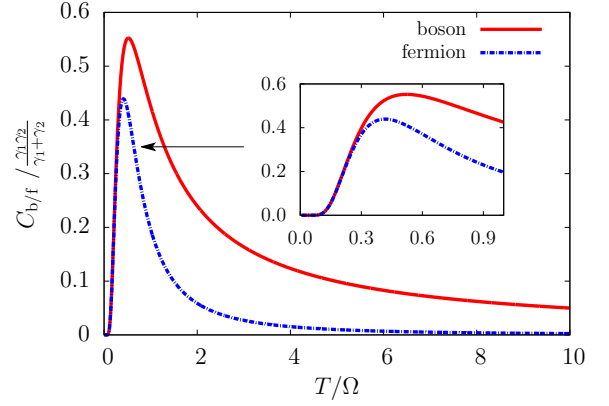}
\par\end{centering}

\protect\caption{\label{fig7}(color online). The ratio $C$ between the refined entropy
production rate $R$ and $(\Delta T/T)^{2}$ vs the temperature $T$.
The red solid line corresponds to the boson case while the blue dashed
line corresponds to the fermion case. Both cases have the same behavior
in the low temperature region. In the high temperature region the
ratio $C$ is proportional to $1/T$ in the boson case, while it is
proportional to $1/T^{2}$ in the fermion case. The peaks of the two
curves correspond to the maximum conductances in boson case and fermion
case respectively. }
\end{figure}

The REPRs for bosonic and fermionic environments can be calculated
directly. The results are
\begin{align}
R_{\mathrm{boson}} & =\frac{\Omega\gamma_{1}\gamma_{2}}{\gamma_{1}(2N_{1}+1)+\gamma_{2}(2N_{2}+1)}\cdot\frac{T_{1}-T_{2}}{T_{1}T_{2}}(N_{1}-N_{2}),\nonumber \\
R_{\mathrm{fermion}} & =\frac{\Omega\gamma_{1}\gamma_{2}}{\gamma_{1}+\gamma_{2}}\cdot\frac{T_{1}-T_{2}}{T_{1}T_{2}}(N_{1}-N_{2}),
\end{align}
where $N_{l}=N(\beta_{l})$. If the two temperatures are nearly equal,
$T_{1}\simeq T_{2}\simeq T$, then we have the estimation
\begin{align}
R_{\mathrm{boson}} & \simeq\frac{\gamma_{1}\gamma_{2}}{\gamma_{1}+\gamma_{2}}\cdot\frac{\Omega^{2}}{T^{2}}\cdot\frac{N(1+N)}{2N+1}\cdot\left(\frac{\Delta T}{T}\right)^{2},\nonumber \\
R_{\mathrm{fermion}} & \simeq\frac{\gamma_{1}\gamma_{2}}{\gamma_{1}+\gamma_{2}}\cdot\frac{\Omega^{2}}{T^{2}}\cdot N(1-N)\cdot\left(\frac{\Delta T}{T}\right)^{2}.
\end{align}
In this case both of the entropy production rates are proportional
to the square of $\Delta T/T$. Thus, the entropy production rate
can be regarded as a response to the ``driving force'' $(\Delta T/T)^{2}$,
and the ratio between the response and the ``driving force'' $C=R/(\Delta T/T)^{2}$
as a ``conductance'' in some sense \cite{Onsager 1930}. In the low
temperature region, the ``conductances'' corresponding to bosonic
and fermionic environments have the same asymptotic behavior:

\begin{equation}
C_{\mathrm{boson/fermion}}\simeq\frac{\gamma_{1}\gamma_{2}}{\gamma_{1}+\gamma_{2}}\cdot\frac{\Omega^{2}}{T^{2}}\mathrm{e}^{-\frac{\Omega}{T}}.
\end{equation}
They both exponentially decays to zero as $T\rightarrow0$. In the
high temperature region, a remarkable difference arises. In this region
we have
\begin{align}
C_{\mathrm{boson}} & \simeq\frac{\gamma_{1}\gamma_{2}}{\gamma_{1}+\gamma_{2}}\cdot\frac{\Omega}{2T},\nonumber \\
C_{\mathrm{fermion}} & \simeq\frac{\gamma_{1}\gamma_{2}}{\gamma_{1}+\gamma_{2}}\cdot\left(\frac{\Omega}{2T}\right)^{2}.
\end{align}
 Thus, the bosonic ``conductance'' $C_{\mathrm{boson}}\propto1/T$
as $T\rightarrow\infty$, while the fermionic ``conductance'' $C_{\mathrm{fermion}}\propto1/T^{2}$
as $T\rightarrow\infty$. Since the ``conductance'' $C$ tends to
zero in both of the limits $T\rightarrow0$ and $T\rightarrow\infty$,
there exists a maximum conductance at a certain finite temperature
$T_{m}$ (see the peak of blue line in Fig.\,\ref{fig7}). The numerical
simulation results for the bosonic and fermionic environments are
presented in Fig.\,\ref{fig7}. It is clearly illustrated that the
conductances, namely, the ratio between REPR and $(\Delta T/T)^{2}$,
are indeed different for bosonic and fermionic environments, especially
in the high temperature region.

\emph{Quantum Coherence Effect on Entropy Production Rate.}--- Now
we consider the factor of quantum coherence. Hereafter, for simplicity
we assume the Markovian property of the dynamics of the quantum open
system and the validity of the rotating wave approximation to the
quantum master equation. Under this assumption, the evolutions of
the diagonal and the off-diagonal parts of the open system's density
matrix are decoupled \cite{Breuer2002}. In other words, ${\cal L}_{l}[\rho_{coh}]$
should have vanishing diagonal elements. Here, $\rho_{coh}$ is the
off-diagonal part of the density matrix, which represents the quantum
coherence of the open system. Thus, $\mathrm{tr}\big[{\cal L}_{l}[\rho_{coh}]\,\ln\rho_{l}^{\mathrm{th}}\big]$
vanishes. This means that the quantum coherence exerts no influence
on the heat flow between the open system and its environment. The
entropy change of the open system $-\mathrm{d}\mathrm{tr}[\rho\ln\rho]/\mathrm{d}t$
can be divided into two parts:
\begin{equation}
-\frac{\mathrm{d}}{\mathrm{d}t}\mathrm{tr}[\rho\ln\rho]=-\mathrm{tr}[\frac{\mathrm{d}\rho_{dia}}{\mathrm{d}t}\ln\rho]-\mathrm{tr}[\frac{\mathrm{d}\rho_{coh}}{\mathrm{d}t}\ln\rho],
\end{equation}
where $\rho_{dia}$ is the diagonal part of the density matrix. Correspondingly,
the REPR in Eq.(\ref{epr_quantum_multi-bath}) can be divided into
two parts:
\begin{eqnarray}
R & = & \left(-\frac{\mathrm{d}}{\mathrm{d}t}\mathrm{tr}[\rho_{dia}\ln\rho]+\sum_{l=1}^{N}\mathrm{tr}\big[{\cal L}_{l}[\rho_{dia}]\,\ln\rho_{l}^{\mathrm{th}}\big]\right)\nonumber \\
 &  & -\mathrm{tr}[\frac{\mathrm{d}\rho_{coh}}{\mathrm{d}t}\ln\rho].
\end{eqnarray}
When $t$ is larger than the time scale of the decoherence, we have
$\rho\simeq\rho_{dia}$ and the term in the first line of the above
equation is none other than the entropy production rate due to the
diagonal part of the quantum open system, which is equal to the classical
REPR inEq.(\ref{REPR}) as we have pointed out before. The term in
the second line can be interpreted as the entropy production rate
due to the evolution of the off-diagonal part of the quantum open
system. Thus we reach the qualitative conclusion that quantum coherence
effect contributes an additional part to the entropy production.

To quantitatively study the effects of quantum coherence, let us concretely
analyze a two-level system coupled with a single heat bath. The Markovian
quantum master equation of this open system reads \cite{Breuer2002}
\begin{eqnarray}
\frac{\mathrm{d}\rho}{\mathrm{d}t} & = & \Gamma_{-}(\sigma_{-}\rho\sigma_{+}-\frac{1}{2}\sigma_{+}\sigma_{-}\rho-\frac{1}{2}\rho\sigma_{+}\sigma_{-})\nonumber \\
 & + & \Gamma_{+}(\sigma_{+}\rho\sigma_{-}-\frac{1}{2}\sigma_{-}\sigma_{+}\rho-\frac{1}{2}\rho\sigma_{-}\sigma_{+}),
\end{eqnarray}
where $\sigma_{+}=\ket{\mathrm{e}}\bra{\mathrm{g}},\ \sigma_{-}=\ket{\mathrm{g}}\bra{\mathrm{e}},$
are the raising and lowering operators of the two-level system respectively.
The evolution of the off-diagonal elements $\rho_{\mathrm{eg}}(t)=\mathrm{exp(-\Gamma t/2)}\rho_{\mathrm{eg}}(0),$
as expected, is decoupled from the diagonal part of the system where
$\Gamma=\Gamma_{+}+\Gamma_{-}$ is the inverse evolution timescale
of the off-diagonal part. Thus the entropy production due to the quantum
coherence effect is
\begin{equation}
R_{con}=-\frac{1}{\alpha(t)}\ln\frac{1+\alpha(t)}{1-\alpha(t)}\frac{\mathrm{d}|\rho_{\mathrm{eg}}|^{2}}{\mathrm{d}t},
\end{equation}
where
\begin{equation}
\alpha(t)=\sqrt{\left(\frac{\Gamma_{-}-\Gamma_{+}}{\Gamma_{-}+\Gamma_{+}}\right)^{2}+4|\rho_{\mathrm{eg}}|^{2}}.
\end{equation}
For a long-time evolution such that $|\rho_{\mathrm{eg}}|\ll(\Gamma_{-}-\Gamma_{+})/\Gamma$,
$R_{con}$ can be estimated as
\begin{equation}
R_{con}\backsimeq\Gamma\beta\Omega\coth\frac{\beta\Omega}{2}\mathrm{e}^{-\Gamma t}|\rho_{\mathrm{eg}}(0)|^{2}.
\end{equation}
It decays exponentially as $t\rightarrow\infty$. 

If initially the diagonal part has already reached its stable value,
the entropy production rate due to the diagonal part of the quantum
open system would remain vaninishing in the evolution. Thus, if we
can ``kick'' an open system, which has been already stabilized to
a thermal state, so that its off-diagonal part becomes non-zero while
its diagonal part remains ``untouched'' , we may be able to observe
the entropy production due to the quantum coherence effect.

\emph{Conclusions and Discussions.}--- In this letter, we try to probe
open quantum systems in contact with two or more baths from entropy
production point of view. We derive a refined formula for the entropy
production rate for such systems. This foumula can well reflect the
effects of statistics and quantum coherence on the entropy production.
In the two-bath case, it turns out that the REPRs in bosonic and fermionic
environments are proportional to the square of temperature difference
$(\Delta T/T)^{2}$ between the two heat baths, but the behaviors
of the so called conductances are quite different in the high temperature
region. This reveals a connection between the entropy production of
the open quantum system and the quantum statistical property of the
baths. The results in this letter are applicable to a non-equilibrium
system weakly coupled to its environment. If the system-bath coupling
is too strong or the coupling spectrum has some exotic structure,
the non-Markovian effects may dominate the long-time behavior of the
systems \cite{Cai 2013}, and the entropy production behavior for
such non-Markovian processes deserves further investigations.

This work was supported by the National Natural Science Foundation
of China (Grant No. 11121403) and the National 973 program (Grant
No. 2012CB922104 and No. 2014CB921403).

\end{document}